\documentclass[preprint,showpacs,preprintnumbers,amsmath,amssymb]{revtex4-1}
 \usepackage{graphicx}% Include figure files
 \usepackage{dcolumn}% Align table columns on decimal point
 \usepackage{bm}% bold math
 \usepackage{ulem}
 \usepackage{enumitem}
 \usepackage{slashed}
\def\q{\bm{q}}
\def\p{\bm{p}}
\def\k{\bm{k}}

\def\r{\bm{r}}
\def\l{\bm{l}}
\def\o{\bm{0}}
\def\vec\epsilon{\bm{\epsilon}}

\def\q{\bm{q}}
\def\p{\bm{p}}
\def\k{\bm{k}}

\def\r{\bm{r}}

\usepackage{xcolor}

\renewcommand\sout{\bgroup \color{red} \ULdepth=-.5ex \ULset}
\begin{document}

\title{Energy loss of heavy quarkonia in hot QCD plasmas}
\author{Juhee Hong}
\affiliation{Department of Physics and Institute of Physics and Applied Physics, Yonsei University,
Seoul 03722, Korea}
\author{Su Houng Lee}
\affiliation{Department of Physics and Institute of Physics and Applied Physics, Yonsei University,
Seoul 03722, Korea}
\date{\today}

\begin{abstract}
We compute the energy loss of heavy quarkonia in high temperature QCD 
plasmas and investigate the energy loss effects on quarkonium suppression. 
Based on the effective vertex derived from the Bethe-Salpeter amplitude 
for quarkonium, the collisional and radiative energy loss are 
determined by quarkonium-gluon elastic scattering and the associated 
gluon-bremsstrahlung, respectively. 
In the energy regime $E< m_\Upsilon^2/T$ the collisional energy loss is 
dominant over the radiative one, and the total energy loss increases with 
the plasma temperature and the initial energy of quarkonium.
Our numerical analysis indicates that 
the medium-induced energy loss of the $\Upsilon$(1S) results in stronger 
suppression at higher momentum, although the energy loss effects are found 
to be small compared with the previous estimates of quarkonium dissociation 
in heavy-ion collisions. 
\end{abstract}

\maketitle

\section{Introduction}

The depletion of high momentum particle production with respect to pp 
collisions signals the formation of a quark-gluon plasma (QGP) in 
heavy-ion collisions. 
Especially, heavy quarks and quarkonia which are mostly formed from the 
initial fusion of partons at an early stage are important probes to 
investigate the transport and thermal properties of the high temperature 
and density matter. 
While quarkonia suppression can be influenced by various mechanisms including 
dissociation and energy loss, the energy loss of heavy quarkonia has not been 
computed and is not described by the heavy quark-antiquark potential.

Energetic particles traversing QCD plasmas suffer energy loss by 
elastic scattering or gluon-bremsstrahlung: drag and diffusion cause 
particles to lose their energies, and incoming high-energy particles can be 
radiatively deprived of fractions of their energies. 
There have been many studies about the energy loss of partons. 
The diffusion processes are dominated by $t$-channel gluon exchange 
with soft momentum transfer which requires hard-thermal-loop resummation 
\cite{braaten-thoma0,braaten-thoma}. 
In the presence of multiple scatterings, destructive interference occurs when 
the formation time is large compared to the 
mean free path \cite{gyulassy-wang,wang1995,bdmps0,bdmps-pT}. 
The radiative energy loss is naively of order $g^2$ higher than the 
collisional one, but there can be enhancement in the limit of soft and 
collinear emission: both radiative and collisional processes of heavy quarks 
can be of the same order in the coupling constant, $-\frac{dE}{dx}\sim g^4T^2$ 
neglecting logarithmic corrections \cite{braaten-thoma,mustafa-thoma}. 
The radiative energy loss is dominant over the collisional one for 
ultrarelativistic partons, whereas the collisional energy loss is not 
negligible for heavy quarks \cite{teaney2005,mustafa2005,djor-coll}.

We are interested in the energy loss of a color singlet 
quarkonium in QCD media such as QGPs or large nuclei. 
After production at an early stage of heavy-ion collisions, quarkonia 
undergo not only dissociation (and regeneration) but also energy loss.
A quarkonium state loses its energy by elastic scattering 
($g+\Upsilon\rightarrow g+\Upsilon$) which can induce gluon radiation. 
Quarkonium diffusion and energy loss have been discussed with potential 
nonrelativistic QCD (pNRQCD) in the regime $E_b\gtrsim T$ \cite{yao-muller}. 
In this work, we will use a formalism developed in our previous 
work \cite{jhong}: for quarkonium dissociation through 
the color-dipole interaction \cite{peskin}, an effective vertex based on the 
Bethe-Salpeter amplitude has been introduced to calculate the next-to-leading 
order cross sections which agree with the results of pNRQCD in the regime 
$T\gtrsim E_b$ \cite{pnrqcd}. 
In the current kinematic range up to $q_T\sim 30$ GeV \cite{cms-PbPb}, 
heavy quarkonia are not ultrarelativistic ($E< m_\Upsilon^2/T$) and thus the 
collisional energy loss can be considerable.

We will discuss how the energy loss of heavy quarkonia can be calculated 
using our formalism of the effective vertex, and estimate the energy loss 
effects on quarkonium spectrum in heavy-ion collisions. 
In Section \ref{coll-eloss}, we calculate the momentum diffusion coefficient 
and the collisional energy loss of weakly bound quarkonia at high temperature.  
In Section \ref{rad-eloss}, we discuss the radiative energy loss by 
gluon-bremsstrahlung associated with quarkonium-gluon elastic scattering. 
Shifting the transverse momentum spectra of the $\Upsilon$(1S) by its mean 
energy loss, we estimate the energy loss effects on the nuclear modification 
factor in Section \ref{raa}.  
Finally, a summary is given in Section \ref{summary}.

\section{Collisional energy loss}
\label{coll-eloss}

\begin{figure}
\includegraphics[width=1.03\textwidth]{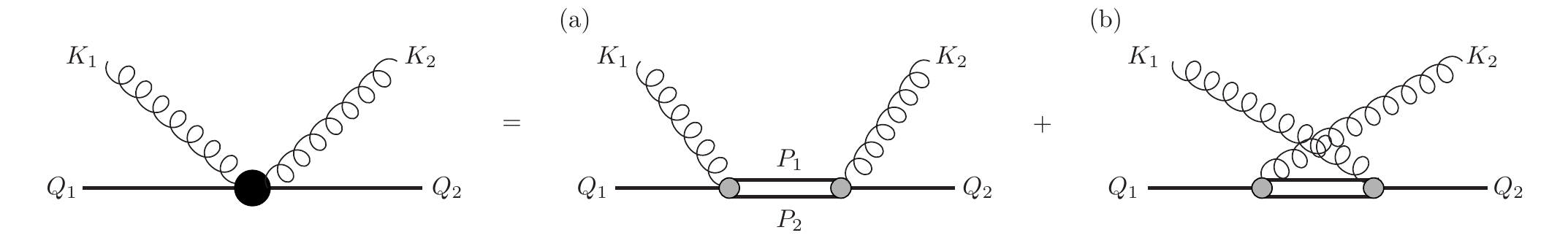}
\caption{
Elastic scattering ($g+\Upsilon\rightarrow g+\Upsilon$) contributing to 
quarkonium diffusion and energy loss.
}
\label{elastic}
\end{figure}

We are interested in diffusion and energy loss of weakly bound quarkonia in 
a high temperature regime where the binding energy is smaller than the 
temperature scale. 
Due to the high melting temperature $T_{\rm melt}\sim 600$ MeV \cite{review}, 
the ground state of bottomonium survives as a color singlet and 
undergoes diffusion in QGP below $T_{\rm melt}$.

To determine the momentum diffusion coefficient of heavy quarkonia, we 
consider quarkonium-gluon elastic scattering. 
In Fig. \ref{elastic} (a) and (b), the dipole interaction of color charge 
with gluon can be described by the following effective vertex derived from the 
Bethe-Salpeter amplitude \cite{song}:
\begin{equation}
\label{vertex}
V^{\mu\nu}(K)=-g\sqrt{\frac{m_\Upsilon}{N_c}}
\left[\k\cdot\frac{\partial\psi(\p)}{\partial \p}\delta^{\mu0}+
k_0\frac{\partial\psi(\p)}{\partial p^i}
\delta^{\mu i}\right]\delta^{\nu j}\frac{1+\gamma^0}{2}\gamma^j
\frac{1-\gamma^0}{2}T^a \, ,
\end{equation}
where $\p=(\p_1-\p_2)/2$ is the relative momentum between heavy quark 
and antiquark.  
With two effective vertices and heavy quark propagators $\Delta(P)$, 
the quarkonium-gluon elastic scattering in Fig. \ref{elastic} (a) has the 
following amplitude \cite{yao-muller}: 
\begin{eqnarray}
\mathcal{M}_{\rm el}^{\mu\nu\rho\sigma(a)}&=& 
\int \frac{d^4P}{(2\pi)^4} \, 
\Delta(P_1)V^{\mu\nu}(K_1)\Delta(P_2)V^{\rho\sigma*}(K_2)  
\, , \nonumber\\
&=&i\frac{g^2m_\Upsilon}{2N_c}\delta^{ab}k_{10}k_{20}
\int\frac{d^3\p}{(2\pi)^3}
\frac{\partial\psi(\p)}{\partial p^i}\delta^{\mu i}
\frac{\partial\psi(\p)}{\partial p^k}\delta^{\rho k}
\frac{\delta^{\nu j}\delta^{\sigma l} \, {\rm Tr}[\gamma^j\gamma^l]}
{2(k_{10}-E_b-\frac{\p^2}{m})} \, ,
\end{eqnarray}
and similarly for Fig. \ref{elastic} (b) except the denominator has 
$2(-k_{20}-E_b-\frac{\p^2}{m})$. 
For $k_{10,20}\gg E_b$, the total matrix element squared 
(averaged over the quarkonium polarization) is 
\begin{equation}
\label{mtx-elt}
|\mathcal{M}_{\rm el}|^2=\frac{32}{81}g^4m_\Upsilon^2
(1+\cos^2\theta_{k_1k_2})
\left[\int\frac{d^3\p}{(2\pi)^3}\left(\frac{\p^2}{m}+E_b\right)|\nabla\psi|^2
\right]^2 \, ,
\end{equation}
where $\theta_{k_1k_2}$ is the angle between 
$\k_1$ and $\k_2$.

The momentum diffusion coefficient is defined by the mean-squared momentum 
transfer per unit time \cite{teaney2005}, 
\begin{multline}
3\kappa=\frac{1}{2q_{10}}\int\frac{d^3\q_2}{(2\pi)^32q_{20}}
\int\frac{d^3\k_1}{(2\pi)^32k_{10}}
\int\frac{d^3\k_2}{(2\pi)^32k_{20}}
n(k_1)[1+n(k_2)]
\\
\times
(2\pi)^4\delta^4(K_1+Q_1-K_2-Q_2)
|\mathcal{M}_{\rm el}|^2(\k_1-\k_2)^2 \, ,
\end{multline}
where $n(k)=1/(e^{k/T}-1)$ is a thermal distribution of gluon.  
Using a Coulombic bound state, 
$|\nabla\psi_{1S}(\p)|^2=2^{10}\pi a_0^7\p^2/[(a_0\p)^2+1]^6$, 
with $a_0^2=1/(mE_b)$ which is satisfied for the Coulombic binding energy, 
we have 
\begin{equation}
\label{kappa}
\kappa=\frac{128\pi g^4T^5}{1215m^2} \, .
\end{equation}
The diffusion coefficient of quarkonium is suppressed by $T^2/m^2$ compared 
with the heavy quark diffusion $\kappa_{\rm HQ}\sim g^4T^3$ \cite{teaney2005}. 
In the QGP temperature region, Eq. (\ref{kappa}) is of the same order of 
magnitude as the results in Ref. \cite{yao-muller} but is much smaller than the 
momentum broadening rate for tightly bound quarkonia in 
Ref. \cite{teaney-erdm}.

In the energy regime $E<m_\Upsilon^2/T$, the collisional energy loss per unit 
length is obtained by the interaction rate of quarkonium weighted by 
$(k_2-k_1)/v$ ($v$ is the quarkonium velocity) 
\cite{braaten-thoma0,braaten-thoma}
\begin{multline}
-\frac{dE}{dx}=\frac{1}{2q_{10}}\int\frac{d^3\q_2}{(2\pi)^32q_{20}}
\int\frac{d^3\k_1}{(2\pi)^32k_{10}}\int\frac{d^3\k_2}{(2\pi)^32k_{20}}
n(k_1)[1+n(k_2)]
\\
\times
(2\pi)^4\delta^4(K_1+Q_1-K_2-Q_2)
|\mathcal{M}_{\rm el}|^2\frac{(k_2-k_1)}{v} \, , 
\end{multline}
where $\cos\theta_{k_1k_2}$ in Eq. (\ref{mtx-elt}) is rewritten in a covariant 
form, 
\begin{equation}
\cos\theta_{k_1k_2}
=1-\frac{(K_1\cdot K_2)m_\Upsilon^2}{(K_1\cdot Q_1)(K_2\cdot Q_1)} \, .
\end{equation}
The $\q_2$ integration is done by the thee-dimensional $\delta$ function, and 
the remaining phase space integral is over 
$k_1, \, k_2, \, \, \theta_{q_1k_1}, \, \theta_{q_1k_2}$, and 
$\phi_{q_1;k_1k_2}$, where $\theta_{q_1k_1}$($\theta_{q_1k_2}$) is the polar 
angle between $\q_1$ and $\k_1$($\k_2$) and $\phi_{q_1;k_1k_2}$ is the 
azimuthal angle between the $\q_1$-$\k_1$ and $\q_1$-$\k_2$ planes. 
We introduce a dummy variable, 
$\int d\omega \, \delta (\omega-k_1+vk_1\cos\theta_{q_1k_1})=1$ 
\cite{teaney2005}, and integrate over the polar angles using the remaining 
$\delta$ functions. 
From $Q_2^2=(Q_1-L)^2$ with the momentum transfer $L$, 
we obtain $Q_1\cdot L=\frac{L^2}{2}$. 
Then 
$k_1-k_2=vk_1\cos\theta_{q_1k_1}-vk_2\cos\theta_{q_1k_2}-\frac{L^2}{2q_{10}}$ 
and the energy conservation yields 
$\delta(k_2-\omega-vk_2\cos\theta_{q_1k_2}-\frac{L^2}{2q_{10}})$, where 
the last term $-\frac{L^2}{2q_{10}}$ is negligible in comparison with the 
other terms of order $T$.
Thus, with  
$\k_1\cdot \k_2=k_1k_2(\cos\theta_{q_1k_1}\cos\theta_{q_1k_2} 
+\sin\theta_{q_1k_1}\sin\theta_{q_1k_2}\cos\phi_{q_1;k_1k_2})$, 
the energy loss is  
\begin{multline}
\label{dEdxeq}
-\frac{dE}{dx}=
\frac{1}{16E(2\pi)^4v^3}\int d\omega
\int_{\frac{\omega}{1+v}}^{\frac{\omega}{1-v}}dk_1
\int_{\frac{\omega}{1+v}}^{\frac{\omega}{1-v}}dk_2 
\int d\phi_{q_1;k_1k_2}\,
n(k_1)[1+n(k_2)]
\\
\times
\frac{(k_2-k_1)}{q_{20}}|\mathcal{M}_{\rm el}|^2
\bigg\vert_{\cos\theta_{q_1k_1}=\frac{k_1-\omega}{k_1v}, \,\, 
\cos\theta_{q_1k_2}=\frac{k_2-\omega}{k_2v}} \, ,
\end{multline}
which can be performed numerically by using the binding energy 
computed in lattice QCD \cite{lattice}.

The effective vertex of Eq. (\ref{vertex}) is based on the dipole interaction 
of color charge with gluon which is valid when the quarkonium size is smaller 
than the inverse energy transfer of the gluon. 
Hence, we consider the quarkonium energy loss at a kinematic regime where 
the temperature and binding energy scales are smaller than 
$\frac{1}{a_0}\sim 1-1.5$ GeV. 
In principle, there can be energy loss of the virtual heavy quarks in Fig. 
\ref{elastic} \cite{yao-muller} and energy loss coming from the color octet 
state prior to quarkonium formation. 
Assuming that the lifetime of the virtual heavy quarks (in the large $N_c$ 
limit) and the color octet state is proportional to the formation time of the 
quarkonium which is related to the inverse of the binding energy, 
the square of the momentum transfer of the virtual heavy quarks and the 
initial color octet state is roughly of order  
$\frac{\kappa_{HQ}}{E_b}\sim \frac{g^4T^3}{E_b}$. 
In the regime $\frac{1}{r}\sim mg^2>T$, $E_b\sim mg^4>g^2T$ and 
the momentum transfer is smaller than the $gT$ scale. 
Since the momentum scale of the heavy quarks is at least of order $T$, we 
ignore the energy loss by the initial color octet state as well as the virtual 
heavy quark diffusion in this work.

\begin{figure}
\includegraphics[width=0.45\textwidth]{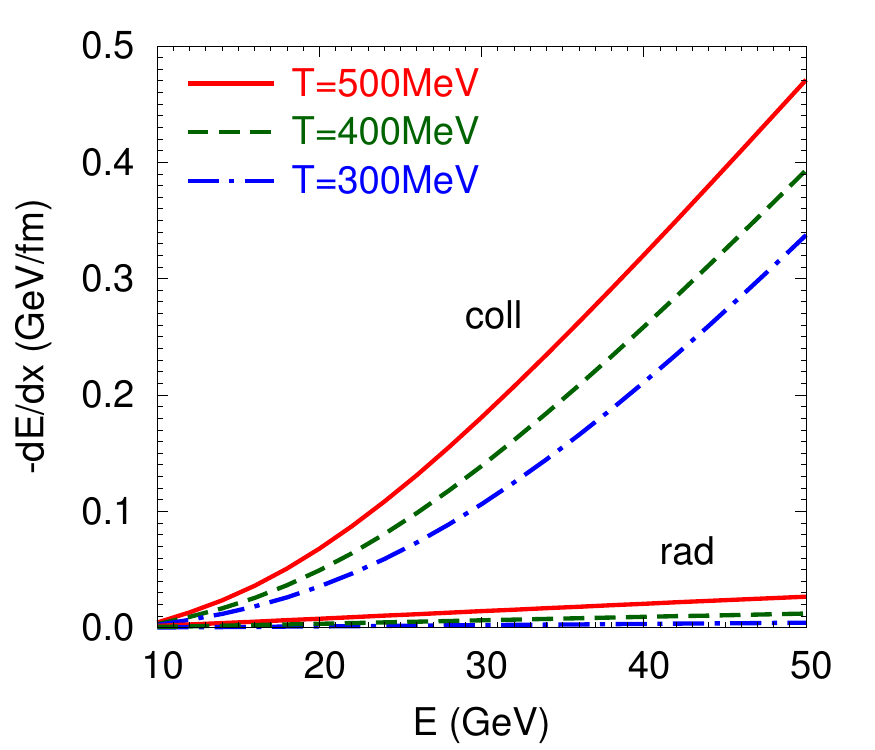}
\caption{
The collisional and radiative energy loss of the $\Upsilon$(1S) 
in a plasma with temperature $T$, as a function of its initial energy. 
$\alpha_s=0.3$ is used. 
}
\label{dEdx}
\end{figure}

Figure \ref{dEdx} shows our numerical results of the $\Upsilon$(1S) energy loss 
with $m=4.8$ GeV, $m_\Upsilon=9.46$ GeV, and $\alpha_s=0.3$. 
The collisional energy loss of the $\Upsilon$(1S) increases with 
its energy and the plasma temperature, and it is less than about half of the 
bottom quark energy loss \cite{braaten-thoma,mustafa-thoma,djor-coll}.

\section{Radiative energy loss}
\label{rad-eloss}

\begin{figure}
\includegraphics[width=0.34\textwidth]{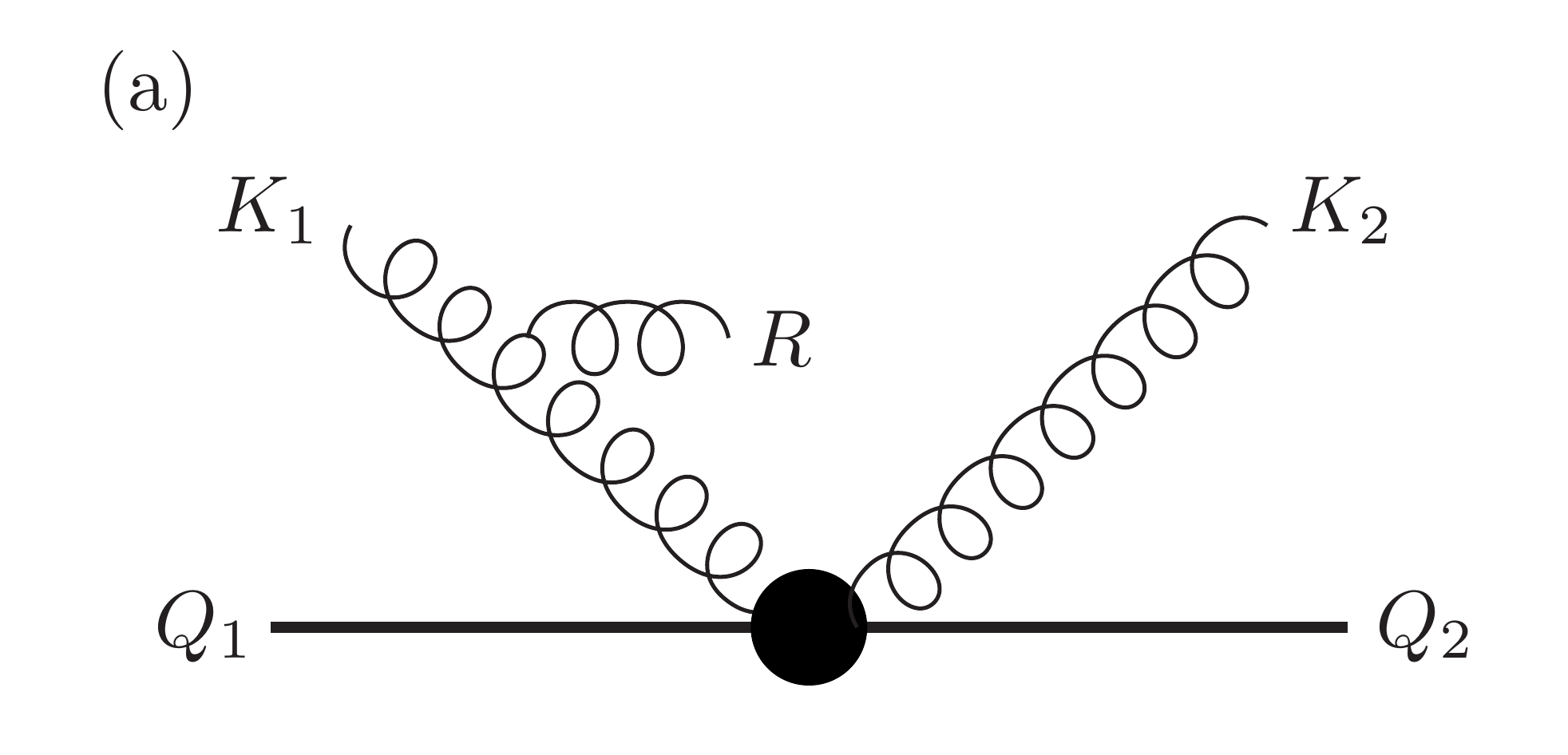}
\qquad \qquad \quad
\includegraphics[width=0.34\textwidth]{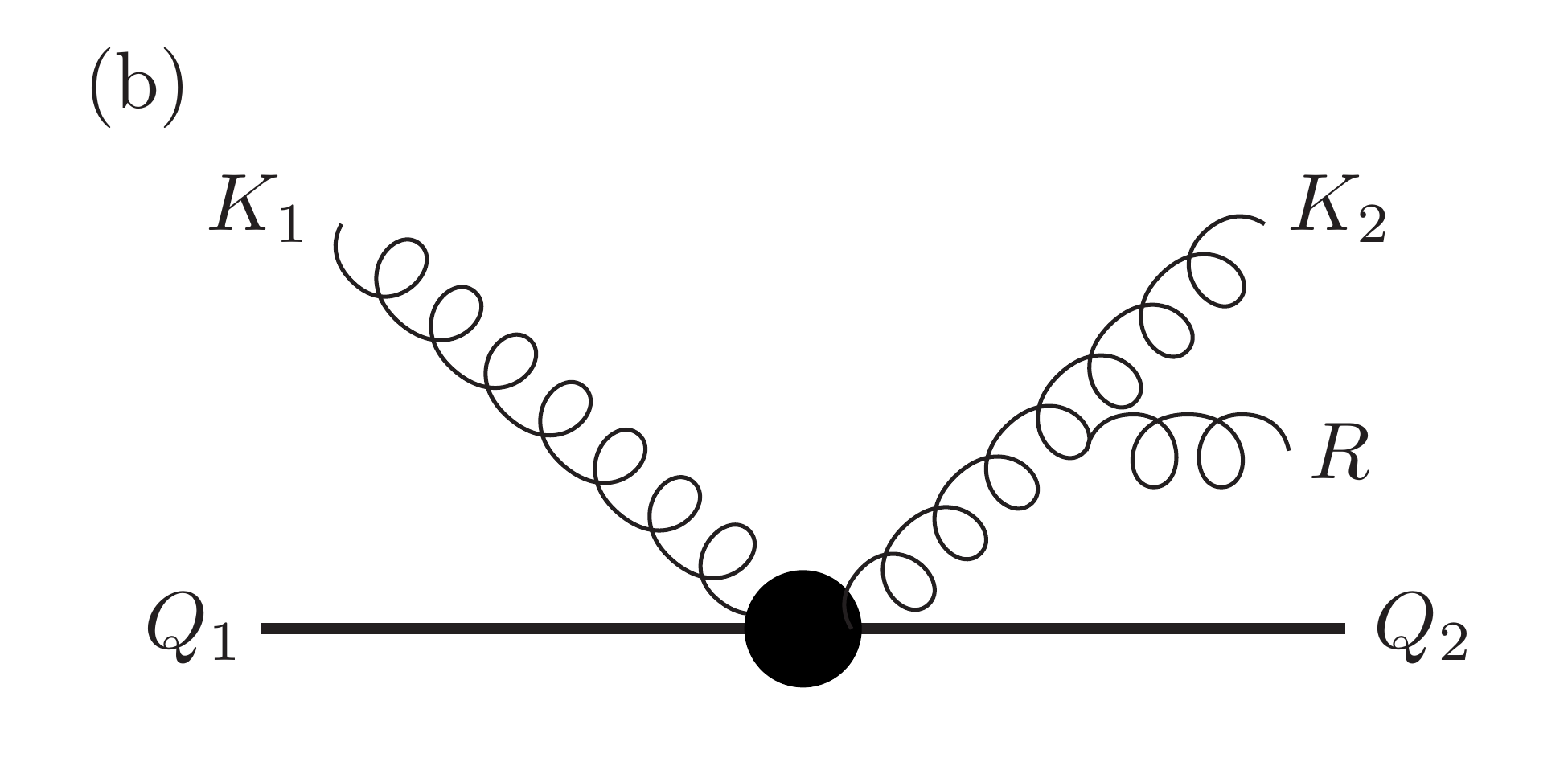}
\caption{
Gluon-bremsstrahlung associated with quarkonium-gluon elastic scattering 
(Fig. \ref{elastic}).
}
\label{fig_rad}
\end{figure}

Quarkonia undergoing quarkonium-gluon elastic scattering induce gluon 
radiation, and the amount of the emitted gluon energy is the radiative energy 
loss. 
In terms of the elastic scattering defined on the left hand side of 
Fig. \ref{elastic}, the lowest order contribution to the energy loss is 
from the processes in Fig. \ref{fig_rad}. 
Gluon emission from heavy quark lines is ignored in the following 
light-cone coordinates \cite{gunion}:
\begin{eqnarray}
\label{coord}
K_1&=&\Big[\sqrt{s}-\frac{m_\Upsilon^2}{\sqrt{s}},0,\o\Big] \, ,
\qquad\qquad 
R=\Big[z\Big(\sqrt{s}-\frac{m_\Upsilon^2}{\sqrt{s}}\Big),
\frac{r_T^2}{z(\sqrt{s}-\frac{m_\Upsilon^2}{\sqrt{s}})},\r_T\Big] \, ,
\nonumber\\
Q_1&=&\Big[\frac{m_\Upsilon^2}{\sqrt{s}},\sqrt{s},\o\Big] \, ,
\qquad\qquad\qquad
L=[l^+, l^-, \l_T] \, ,
\end{eqnarray}
where $z$ is the momentum fraction of the emitted gluon relative to the 
maximum available, and the components of the momentum transfer $L$ are 
determined by the on-shell conditions, $K_2^2=(K_1+L-R)^2=0$ and 
$Q_2^2=(Q_1-L)^2=m_\Upsilon^2$. 
In the gauge $A^+=0$, the polarization of the radiated gluon is specified by 
$\epsilon\cdot R=0$ and $\epsilon^+=0$, 
\begin{equation}
\epsilon=\Big[0,\frac{2\vec\epsilon_T\cdot \r_T}{z(\sqrt{s}
-\frac{m_\Upsilon^2}{\sqrt{s}})},\vec\epsilon_T\Big] \, .
\end{equation}
For soft gluon emission, the radiation processes are 
factorized into elastic scattering and gluon emission whose amplitude is 
the emitted gluon field multiplied by an additional gluon propagator. 
In the limit $r_T\gg l_T$, the leading contribution is as follows: 
\begin{equation}
\mathcal{M}_{\rm rad}=2gCz\frac{\vec\epsilon_T\cdot\r_T}{\r_T^2} \, 
\mathcal{M}_{\rm el} \, ,
\end{equation}
where $C$ is the color factor associated with Fig. \ref{fig_rad}, divided by 
the factor in the absence of radiation ($C^2=3$).  
The energy of the emitted gluon is small compared to that of the parent gluon 
for soft radiation, but it is still assumed to be larger than its transverse 
momentum.

The radiative energy loss associated with quarkonium-gluon elastic 
scattering can be determined by the emitted gluon spectrum, 
\begin{equation}
\label{gspec}
\int dn_g
=\int \frac{d^4R}{(2\pi)^4}\, 2\pi\delta (R^2)
\left\vert\frac{\mathcal{M}_{\rm rad}}{\mathcal{M}_{\rm el}}\right\vert^2
=\int\frac{dz}{z}\int \frac{d^2\r_T}{(2\pi)^3}\, 6g^2 \frac{z^2}{\r_T^2} \, .
\end{equation}
The mean energy loss is the average of the probability of emitting a gluon 
times the gluon energy, 
\begin{eqnarray}
\label{delE}
\delta E
=\int dn_g \, r_0
=\frac{g^2}{4\pi^2}\Big(\sqrt{s}-\frac{m_\Upsilon^2}{\sqrt{s}}\Big) 
\ln\bigg[\frac{\sqrt{s}-\frac{m_\Upsilon^2}{\sqrt{s}}}{2m_D}\bigg]
\, ,
\end{eqnarray}
where $r_0$ is the energy of the emitted gluon. 
In Eq. (\ref{gspec}), the maximum transverse momentum of the gluon is given by 
$zk_1=z(\sqrt{s}-\frac{m_\Upsilon^2}{\sqrt{s}})/2$, and the Debye screening 
mass $m_D\sim gT$ has been chosen for the minimum: our results and the 
following discussion are not very sensitive to variation in the limits. 
For an estimate, we use $s\simeq m_\Upsilon^2+2Ek_1$ with a mean thermal 
energy $k_1\sim 3T$.   
The average radiative energy loss per unit length is estimated by 
$-dE/dx\approx\delta E/\lambda$ with the wavelength 
$\lambda=1/(\sigma_{\rm el}\, \rho)$, where $\sigma_{\rm el}$ 
is the cross section of quarkonium-gluon elastic scattering in 
Fig. \ref{elastic} and 
$\rho=16\int\frac{d^3\k}{(2\pi)^3}n(k)=\frac{16\zeta(3)T^3}{\pi^2}$ is the 
gluon density.

We present the $\Upsilon$(1S) radiative energy loss in Fig. \ref{dEdx}, 
comparing with the collisional energy loss. 
The $\Upsilon$(1S) radiatively loses more energy in a hotter medium, but  
the effect is much smaller than the collisional energy loss which grows more 
rapidly at high energy. 
In comparison with the radiative energy loss of a bottom quark 
\cite{djor-rad}, the energy loss of the $\Upsilon$(1S) is approximately an 
order of magnitude smaller at least.

\section{Nuclear modification factor}
\label{raa}

In the previous two sections, we have computed the collisional and 
radiative energy loss of the $\Upsilon$(1S). 
Since the energy loss depends on the medium temperature, it changes with time 
as the plasma expands and cools down in heavy-ion collisions. 
In this section, we use the medium-induced energy loss to calculate the 
total energy loss during evolution, and then investigate the energy loss 
effects on nuclear modification factors in the central rapidity region.

\begin{figure}
\includegraphics[width=0.45\textwidth]{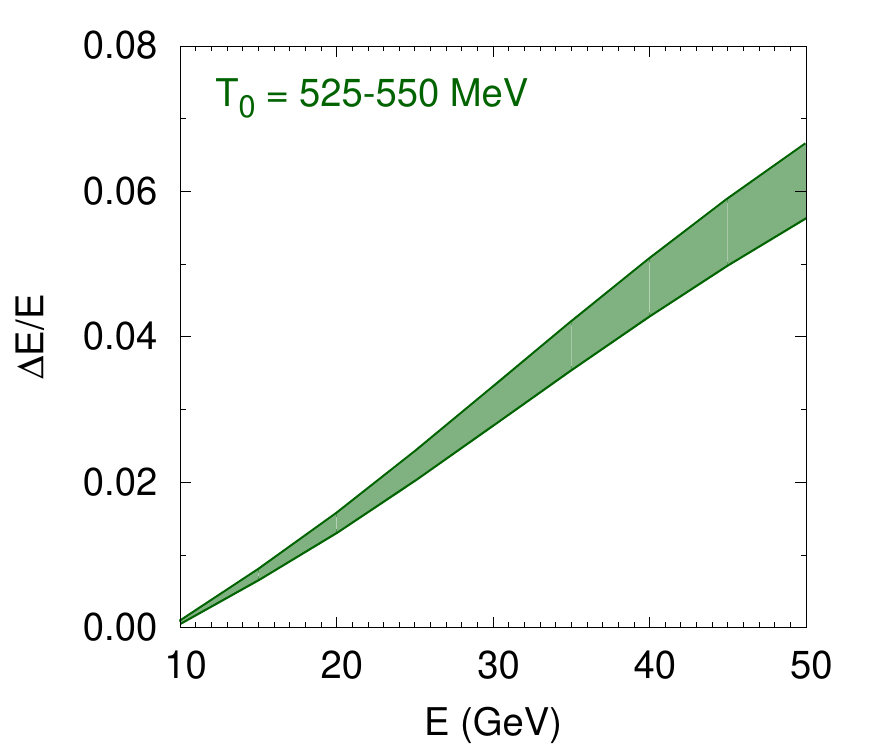}
\caption{
The fractional energy loss $\frac{\Delta E}{E}$ of the $\Upsilon$(1S) 
for a Bjorken expansion with the initial temperature $T_0=525-550$ MeV, 
the higher temperature being the upper boundary.
}
\label{dEE}
\end{figure}

Quarkonium production has not been fully understood and the theoretical 
prediction involves large uncertainties even in the absence of a nuclear 
medium. 
For these reasons, we exploit quarkonium momentum spectra measured in pp 
collisions and then estimate the energy loss effects in AA collisions.
The transverse momentum spectrum of the $\Upsilon$(1S) in pp collisions can be 
parameterized as 
$\frac{d\sigma_{\rm pp}}{d^2\q_T}\propto\frac{1}
{[(\q_T/\Lambda)^2+1]^{\alpha}}$ 
($\Lambda=6.05$ GeV and $\alpha=2.44$) from the data in Ref. \cite{cms-PbPb}. 
Using the momentum spectrum as an initial unquenched one, 
the energy loss effects can be realized approximately by a shift of the 
momentum spectrum by the $\Upsilon$(1S) energy loss. 
For an expanding plasma undergoing a Bjorken expansion \cite{bjorken}, 
$T(t)=T_0\,(\frac{t_0}{t})^{1/3}$, the total energy loss during the evolution 
is $\Delta E=-\int_{t_0}^{t_f}dt \, v \frac{dE}{dx}$, where $t_0\sim 0.3$ fm/c 
and $t_f\sim 7$ fm/c at the phase transition 
(as assumed in Ref. \cite{jhong2}). 
If the energy loss is small ($\frac{\Delta E}{E}\ll 1$, see Fig. \ref{dEE}), 
the effects on normalized transverse momentum spectra can be approximated as 
\cite{bdms-quenching}
\begin{equation}
\frac{d\sigma_{\rm AA}(E)}{d^2\q_T}=
\frac{d\sigma_{\rm pp}(E+\Delta E)}{d^2\q_T} \, .
\end{equation}
Then we estimate the nuclear modification factor by the energy loss effects,
\begin{equation}
R_{\rm AA}(q_T)
=\frac{\, \, \, \frac{d\sigma_{\rm AA}(E)}{d^2\q_T} \,\, \, } 
{\frac{d\sigma_{\rm pp}(E)}{d^2\q_T}} \, .
\end{equation}

\begin{figure}
\includegraphics[width=0.45\textwidth]{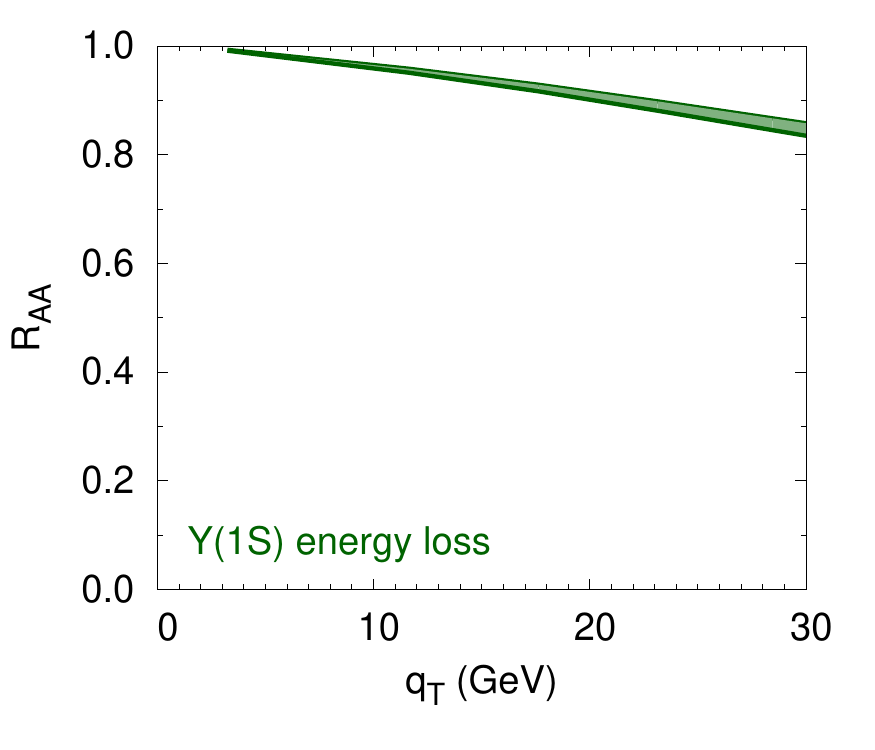}
\caption{
The energy loss effects on the $\Upsilon$(1S) $R_{\rm AA}$ factor in an 
expanding plasma with the initial temperature $T_0=525-550$ MeV (as assumed 
in Ref. \cite{jhong2}) for heavy-ion collisions at 
$\sqrt{s_{NN}}=2.76,\,5.02$ TeV. 
}
\label{Raa}
\end{figure}

In Ref. \cite{jhong2} we have computed the nuclear modification factor of 
the $\Upsilon$(1S) by dissociation and regeneration in PbPb collisions 
at $\sqrt{s_{NN}}=2.76,\,5.02$ TeV, and found that  
the numerical results depend on initial conditions with 
significant uncertainties at an early stage when 
quarkonia formation is in progress. 
The quarkonium energy loss can also affect the initial spectrum at the initial 
time QGP is formed (as well as during a Bjorken expansion): the energy loss 
effects at the early stage might be important. 
Using the same initial conditions as Ref. \cite{jhong2} without the energy 
loss at the beginning, Fig. \ref{Raa} shows the $\Upsilon$(1S) 
$R_{\rm AA}$ factor determined by its energy loss during time evolution.  
Because the collisional energy loss increases with momentum as seen in 
Fig. \ref{dEdx}, the $\Upsilon$(1S) is more suppressed at larger momentum. 
For higher initial temperature, we obtain stronger suppression.

The hot-medium effects can be obtained by including both quarkonium energy 
loss and dissociation-regeneration.  
In an effective field theory framework, dissociation and recombination 
have been systematically studied in Refs. \cite{yao1,yao2,yao3,yao4}.  
Noting that the enhancement by regeneration tends to grow with momentum 
in Ref. \cite{jhong2}, 
the energy loss as in Fig. \ref{Raa} can weaken the enhancement influence 
at high momentum to be more consistent with the data.   
As the regeneration effects are more pronounced with softer initial 
distributions of quarkonia and heavy quarks, the energy loss effects become 
more significant when the $\Upsilon$(1S) spectrum is softer. 
Although the energy loss makes a little contribution compared with the almost 
momentum-independent suppression $R_{\rm AA}\sim0.4$ measured at the LHC 
\cite{cms-PbPb,cms-PbPb2}, 
the $\Upsilon$(1S) energy loss can affect the high momentum spectra 
up to $\sim 15\%$ at $q_T\sim 30$ GeV, and thus 
the diffusion and energy loss of heavy quarkonia, 
together with dissociation and regeneration, need to be taken into account to 
analyze the experimental data.

For feed-down, we can apply the calculation of 
quarkonium energy loss to the excited states of bottomonia as well. 
Using a Coulombic bound state, 
$|\nabla\psi_{2S}(\p)|^2=2^{7}\pi a_0^7\p^2[(a_0\p)^2-\frac{1}{2}]^2/[(a_0\p)^2+\frac{1}{4}]^8$,
with a quarter of the binding energy of the 
ground state, the energy loss of the $\Upsilon$(2S) is more than six times 
as large as the $\Upsilon$(1S) energy loss. 
Larger energy loss as well as thermal width of excited states might lead to 
sequential suppression of bottomonia.

\section{Summary}
\label{summary}

We have presented the first estimate of quarkonium energy loss in hot QCD 
plasmas using an effective vertex between quarkonium and gluon. 
Based on our formalism derived from the Bethe-Salpeter amplitude, we have 
investigated the leading effects on the transverse momentum spectra of the 
$\Upsilon$(1S). 
The collisional energy loss is obtained by convoluting the interaction rate of 
quarkonium-gluon elastic scattering and the energy transfer. 
The radiative energy loss is realized through gluon-bremsstrahlung associated 
with the elastic scattering and is estimated by a convolution of the 
probability of emitting a gluon and its energy. 
Our numerical analysis indicates that the collisional energy loss is 
dominant over the radiative one in the energy regime $E< m_\Upsilon^2/T$, 
and increases with the quarkonium energy and the plasma temperature. 
In comparison with the bottom quark energy loss, the energy loss of the 
$\Upsilon$(1S) is smaller by a factor of $2$ at least.

The energy loss of heavy quarkonia affects their transverse momentum spectra 
and is reflected on the nuclear modification factors, resulting in quarkonia 
suppression. 
Adopting a simple shift of the momentum spectra by the energy loss for a 
Bjorken expansion, we have estimated the energy loss effects on the 
$\Upsilon$(1S) $R_{\rm AA}$ factor.  
The energy loss is found to provide a small effect on the $\Upsilon$(1S) 
suppression compared with quarkonium dissociation, but it is 
necessary to consider the hot-medium effects and to analyze the 
experimental data in heavy-ion collisions.  
Furthermore, the energy loss of heavy quarkonia might be partially responsible 
for sequential suppression even in small systems.  
As the medium-induced energy loss is sensitive to many factors including 
path-length and geometrical effects in nuclear collisions, the mean energy 
loss is not sufficient and more systematic studies need to be undertaken.

\section*{Acknowledgments}

We would like to thank Yongsun Kim for useful discussions. 
This work is supported by the National Research Foundation of Korea
(NRF) grant funded by the Korea government (MSIT) (No. 2018R1C1B6008119).

\end{document}